\documentclass[submission,copyright,creativecommons]{eptcs}
\providecommand{\event}{Refinement Workshop 2018} 
\usepackage{breakurl}             
\usepackage{underscore}           

\usepackage{times}
\usepackage{hyperref}
\usepackage{xcolor}
\usepackage{amssymb}

\usepackage{laws}

\usepackage{imakeidx}
\makeindex
\indexsetup{noclearpage}

\newcommand{\draftonly}[1]{}

\newcommand{\enqueue}{write}
\newcommand{\dequeue}{read}

\definecolor{darkgreen}{rgb}{0,0.6,0}

\newcommand{\implies}{\mathrel{\Rightarrow}}
\newcommand{\sdefs}{\mathrel{\widehat{=}}}
\newcommand{\refsto}{\mathrel{\sqsubseteq}}

\newcommand{\cat}{\mathbin{\raise 0.8ex\hbox{$\frown$}}}
\newcommand{\prefix}{\mathrel{\Keyword{prefixof}}}
\newcommand{\suffix}{\mathrel{\Keyword{suffixof}}}
\newcommand{\seq}{\Keyword{seq}}

\newcommand\eventually{\mathop{\Diamond}}
\newcommand{\always}{\mathop{\Box}}

\newcommand{\figurerule}{\rule{\textwidth}{0.5pt}}

\newcommand{\Keyword}[1]{\mathop{\mathbf{#1}}}
\newcommand{\Terminate}{\Keyword{terminate}}
\newcommand{\Encode}{\Keyword{encode}}
\newcommand{\Await}{\Keyword{await}}

\newcommand{\Do}{\mathop{\Keyword{do}}}
\newcommand{\Guar}{\mathop{\Keyword{guar}}}

\newcommand{\Initially}{\Keyword{initially}}
\newcommand{\Invariant}{\Keyword{invariant}}

\newcommand{\Od}{{\Keyword{od}}}
\newcommand{\Rely}{{\Keyword{rely}}}

\newcommand{\Resource}{\Keyword{resource}}

\newcommand{\Term}{\Keyword{term}}

\newcommand{\Where}{\Keyword{where}}
\newcommand{\With}{\Keyword{with}}

\newcommand{\with}[2]{\With\,#1\,\Do\,#2\,\Od}
\newcommand{\withawait}[3]{\With\,#1\,\Await\,#2\,\Do\,#3\,\Od}

\newcommand{\Pre}[1]{\{#1\}}
\makeatletter
\newcommand{\Frame}[2]{\ifx\@empty#1\else#1\!:\!\fi#2}
\def\Spec{\@ifnextchar*{\@Spec}{\@@Spec}}
\def\@Spec*#1#2#3{\ifx\@empty#1\else#1\colon\fi
   [{#2}\ifx\@empty#2\else,~\fi#3]}
\def\@@Spec#1#2#3{\ifx\@empty#1\else
   \begin{array}{@{}l@{}}#1\colon\end{array}\!\!\fi%
   \left[{\begin{array}{@{}l@{}}#2\end{array}}\ifx\@empty#2\else~,~~\fi
   \begin{array}{@{}l@{}}#3\end{array}\right]}
\makeatother

\newcommand{\Atomic}[1]{\langle#1\rangle}
\newcommand{\EnvAtomic}[1]{\epsilon\Atomic{#1}}

\newcommand{\Comment}[1]{~~~~\mbox{\color{darkgreen}-- #1}}
\newcommand{\Const}[1]{\mathsf{#1}}
\newcommand{\True}{\Const{true}}
\newcommand{\False}{\Const{false}}
\newcommand{\rely}[1]{\Rely~#1}
\newcommand{\guar}[1]{\Guar~#1}

\newcommand{\Seq}{\mathbin{\mathbin{;}}} 

\newcommand{\Om}[1]{#1^{\omega}}
\newcommand{\nondet}{\mathbin{\sqcap}}
\newcommand{\together}{\mathbin{\Cap}}

\makeatletter
\newcommand{\ChainRel}[1]{\crcr \noalign{\penalty\interdisplaylinepenalty}
  \hspace*{-1em}#1
  \@ifnextchar*{\@ChainRelCommment}{}}
\newcommand{\ChainRelFormat}[1]{\mbox{\color{blue}~~~#1}}
\def\@ChainRelCommment*[#1]{\ChainRelFormat{#1}
  \crcr \noalign{\penalty\interdisplaylinepenalty}}
\newcommand{\StartRef}[1]{\hspace*{-1.5em}(\ref{#1}) \refsto
  \@ifnextchar[{\@StartRefCommment}{}}
\def\@StartRefCommment[#1]{\mbox{#1}
  \crcr \noalign{\penalty\interdisplaylinepenalty}}
\makeatother

\newcommand{\entails}{\mathrel{\Rrightarrow}}

\newcommand{\Refsto}{\ChainRel{\refsto}}

\newcommand{\Equals}{\ChainRel{=}}

\DeclareSymbolFont{italics}{\encodingdefault}{\rmdefault}{m}{it}
\makeatletter
\def\@setmcodes#1#2#3{{\count0=#1 \count1=#3
    \loop \global\mathcode\count0=\count1 \ifnum \count0<#2
    \advance\count0 by1 \advance\count1 by1 \repeat}}
\@setmcodes{`A}{`Z}{"7\hexnumber@\symitalics41}
\@setmcodes{`a}{`z}{"7\hexnumber@\symitalics61}
\makeatother

\usepackage[colorinlistoftodos]{todonotes}

\newcounter{hours}
\newcounter{minutes}
\newcommand{\printtime}{%
  \ifthenelse{\value{hours}<10}{0}{}\thehours:%
  \ifthenelse{\value{minutes}<10}{0}{}\theminutes}
\newbox{\time}
\savebox{\time}{\printtime}

\begin{document}

\title{Some Challenges of Specifying \\Concurrent Program Components%
}
\author{Ian J. Hayes
\draftonly{\fbox{\today}}
\institute{School of Information Technology and Electrical Engineering, \\ 
The University of Queensland, Australia
  \draftonly{\\\vspace*{2ex} \today~\printtime}
}}
\def\titlerunning{Some Challenges of Specifying Concurrent Program Components \draftonly{(\today~\usebox{\time})}}
\def\authorrunning{Ian J. Hayes \draftonly{(\today~\usebox{\time})}}
\maketitle

\begin{abstract}
The purpose of this paper is to address some of the challenges of 
formally specifying components of shared-memory concurrent programs.
The focus is to provide an abstract specification of a component
that is suitable for use both by clients of the component and
as a starting point for refinement to an implementation of the component.
We present some approaches to devising specifications, 
investigating different forms suitable for different contexts.
We examine handling atomicity of access to data structures,
blocking operations and progress properties, 
and 
transactional operations that may fail and need to be retried.
\end{abstract}

\section{Introduction}

The objective of this paper is to address challenges to do with specifying concurrent program components
in order to promote discussion about alternative approaches.
Our main foci are atomicity, blocking operations and transactional operations
in the context of rely/guarantee specifications \cite{Jones83a,Jones83b}.
Our aim is to present the ideas rather than a fully formal development.
Specifications play an important role in decoupling the use of a component from its detailed implementation.
Often the role of specifications as a starting point for refinement to an implementation is emphasised
but here we would like to balance that with their role of being used by other components.
Hence we focus on a top-down approach to concurrent program specification,
rather than a bottom-up approach.

\paragraph{Sequential programs.}

For sequential programs conventional, Floyd/Hoare-style specifications \cite{Floyd67,Hoare69a}
in terms of \emph{preconditions} and \emph{postconditions} form the basis of component specifications,
however, pre and post conditions alone are inadequate for specifying concurrent operations 
because they do not handle interference between the operations.

\paragraph{Shared variable concurrency.}

First, to state the obvious, variables that are local to a thread are not subject to interference
and hence can be treated in a manner similar to their use in a sequential program.
For variables shared between parallel threads, interference becomes an issue.
An important consideration is whether access (e.g.\ read or write) 
to variables is \emph{atomic} or not.
At the lowest level, atomicity is determined by the machine hardware and
properties like its atomic access ``word'' size.
A further complication at the hardware level is that, due to caches and write buffers,
the order of write/read accesses to shared memory may not respect the sequential order of instruction execution.
But perhaps we get a bit ahead of ourselves if we worry about these issues when considering specifications.

A thread that needs to perform multiple atomic accesses can be subject to \emph{data races}
where variables are updated in parallel by concurrent threads 
and the thread may see inconsistent data.
At a more abstract level, operations may be required to be atomic with respect to a shared data structure.
The implementation of such operations may require locks to ensure sequentialisation of access to the data structure
or it may use more sophisticated non-blocking algorithms that achieve the effect of operation atomicity 
by utilising hardware-level atomicity~\cite{Scott13}.

A concept commonly used to show an implementation is valid is \emph{linearizability} \cite{HW90},
whereby parallel execution of a set of operations on a shared data structure is considered valid
if it is equivalent to some linear (sequential) execution of the same operations
(subject to certain requirements).

\paragraph{Verifying concurrent programs.}

Early rules for reasoning about parallel programs by Hoare~\cite{Hoare75}
utilised preconditions and postconditions 
but had strict disjointness requirements on program variables occurring in parallel threads,
which effectively ruled out interference between parallel threads.
The approach of Owicki and Gries \cite{Owicki75,OwickiGries76,OwickiGries-CACM-76} 
treats parallel components like sequential programs
with intermediate assertions between each atomic step
but then requires an extensive interference-freedom proof.
Concurrent separation logic \cite{BrookesSepLogic07,OHearn07} also leverages disjointness
but does so in a more fine-grained and dynamic manner because it handles assertions 
over objects in the heap.

An early compositional approach to handle interleaved interference on shared variables was 
the rely/guarantee approach of Jones \cite{Jones81d,Jones83a,Jones83b}.
Jones extended pre/post specifications with a \emph{rely} condition,
a binary relation between program states expressing an assumption about 
the tolerable interference that any step of the environment of the thread can impose on the shared variables.
To constrain the interference generated by a thread, Jones uses a \emph{guarantee} condition,
also a binary relation on states limiting the changes a program step of a thread can make to the shared variables.
The guarantee is required to be reflexive (i.e.~it contains the identity relation)
so that the program may make stuttering steps that do not change the observable variables.
The rely of each thread must be implied by the guarantees of all threads that run in parallel with it.
The rely/guarantee approach does not dictate any particular granularity of atomicity,
however, it does require all steps of a thread to satisfy its guarantee as long as
all steps of the thread's environment satisfy its rely.

\paragraph{Blocking.}

Operations may block waiting for ``communication'' from another thread.
For example, an operation wanting to read a message from a communication channel 
may need to wait for a message to be written to the channel by another thread.
Specifications of such operations need to be able to express such \emph{waiting} criteria.
This affects the termination behaviour of the operation.
For example, if a message is never written to a channel,
a read from the channel will never terminate (block forever).
We consider two approaches to specifying such operations.
\begin{itemize}
\item
Using an explicit $\Await$ construct that allows both nonterminating behaviour if the $\Await$ blocks forever
and terminating behaviour if it becomes unblocked.
\item
An implicit approach that specifies under what conditions an operation is guaranteed to terminate
as well as its behaviour when it does terminate,
e.g.\ a read on a message channel is guaranteed to terminate if the channel is non-empty.
\end{itemize}
These forms can give equivalent specifications,
for which the explicit form may be more useful for refining the operation,
while the implicit form makes it easier to reason about using the operation.
Another source of blocking is at the implementation level
where atomicity constraints on operations may lead to the use of locks
that lead to an implementation blocking awaiting a lock.

\paragraph{Transactional operations.}

Utilising locks can generate bottlenecks because operations requiring the locks are sequentialised,
and on multi-processor architectures they also generate more costly memory synchronisation primitives.
One approach to avoiding (or minimising) locks is to implement operations that do most of their work locally
and then perform a final atomic commit step that may fail if another operation has committed 
while the first operation was executing based on the old data~\cite{Scott13}.
Such implementations may suffer from starvation
where they can repeatedly try and fail, potentially forever
if there is continual interference from competing parallel operations.
Note that in these situations, one of the competing threads may succeed
but an individual thread may be pre-empted every time and never succeed.
Fair scheduling is assumed, i.e. every thread is executed eventually,
but that does not prevent an operation that a thread is attempting to execute 
from failing due to interference and needing to be retried.

Specifying such operations has to allow for the case in which an operation is continually thwarted
and may never terminate, while guaranteeing termination if the interference on the data structure eventually quiesces.
Using $\Om{c}$ to represent the execution of a command $c$ zero or more times,
including possibly infinitely many times,
such specifications have the general form
\begin{equation}
  \Om{fail} \Seq succeed
\end{equation}
where $fail$ represents the operation failing due to interference
(and not changing the state)
and $succeed$ represents the operation successfully completing
(and updating the state once).
The iteration $\Om{fail}$ may execute $fail$ infinitely many times 
representing the continual thwarting by interference from parallel operations.
Arguments about termination usually need to resort to either 
\begin{itemize}
\item
arguments that interference eventually quiesces altogether,
\item
timing arguments based on minimal separation between operations in any single thread
leading to a situation in which interference will eventually quiesce for long enough for the operation to succeed,
and
\item
arguments based on probabilities of two (or more) operations overlapping and competing.
\end{itemize}
Note that probability bounds can be derived from timing bounds.
The probability can be sensitive to load, 
i.e.\ the more threads competing, the lower the probability of success of any single operation.
And probabilities can also be sensitive to the execution time of an operation:
the longer it executes, the more likely it is to overlap with a competing operation.
Of course, such timing and probability arguments depend on the context of the use of the data structure
and can be tricky in practice.

Implicit specifications can also be used for such operations 
by specifying the conditions under which they are guaranteed to terminate.

\paragraph{Overview.}

Sect.~\ref{S-atomicity} addresses specifying atomic operations on a shared data structure (or resource).
It examines the use of Hoare's $\With$ statement \cite{Hoare71g} and 
how it interacts with rely and guarantee conditions.
Sect.~\ref{S-blocking} examines blocking operations giving both explicit waiting conditions
and more implicit specifications using a temporal logic formula under which an operation terminates.
Sect.~\ref{S-transactional} looks at transactional operations 
that may either succeed, or try and fail, possibly indefinitely.
Both explicit waiting and implicit temporal logic specifications are considered.

\pagebreak[3]

\section{Specifying atomicity}\label{S-atomicity}

As an example, consider a message queue with operations to enqueue and dequeue messages.
If there are separate concurrent threads enqueuing and dequeuing,
each operation needs to (appear to be) be atomic,
i.e.\ other operations cannot observe the state part way through the operation.
If the queue were used in a purely sequential program, 
the $\enqueue$ and $\dequeue$ operations can be specified by 
Morgan-style specification commands \cite{TSS,Morgan94} as follows.
\begin{eqnarray*}
  \enqueue(v : Val) & \sdefs & \Spec{qu}{}{qu' = qu \cat [v]} \\
  \dequeue() res : Val & \sdefs & \Pre{qu \neq [~]} \Seq \Spec{res,qu}{}{qu = [res'] \cat qu'}
\end{eqnarray*}
The $\enqueue$ operation takes a value $v$ to append to the queue.
Its postcondition is $qu' = qu \cat [v]$,
in which $qu'$ stands for the final value of the queue,
$qu$ stands for the initial value,
``$\cat$'' is sequence concatenation, and
$[v]$ is the singleton sequence containing $v$.
The $\enqueue$ operation modifies only the queue and hence it has a frame of $qu$ (before the colon).
The $\dequeue$ operation returns a value $res$ that is the head of the queue
and removes it from the queue in the process.
It has a precondition that the queue is non-empty, which is written as an assertion command (in braces).

To extend the operation specification to handle concurrency, as presented in Fig.~\ref{F-queue},
the operations need to be augmented with rely and guarantee conditions
and the atomicity of the operations needs to be handled.
In Fig.~\ref{F-queue} the relies and guarantees are represented as 
rely and guarantee commands~\cite{DaSMfaWSLwC,FM2016atomicSteps,FMJournalAtomicSteps}.
The guarantee command $(\guar{g})$ restricts every atomic program step of the thread to satisfy $g$.
The rely command $(\rely{r})$ represents an assumption that 
every environment step satisfies $r$;
if an execution trace performs an environment step not satisfying $r$, 
any behaviour whatsoever is allowed from that point 
(i.e.~it aborts in a manner similar to the precondition command $\Pre{p}$ aborting 
if the initial state does not satisfy $p$).
The rely and guarantee commands are combined with 
the remainder of the specification using 
the weak conjunction operator ``$\together$''~\cite{HayesJonesColvin14TR,AFfGRGRACP}.
The weak conjunction $c_0 \together c_1$ performs steps allowed by both $c_0$ and $c_1$
unless either $c_0$ or $c_1$ aborts at some point, in which case their weak conjunction aborts from that point.
Weak conjunction is a specification operator, rather than a programming operator.

For the message queue we assume there is a single writer thread performing $\enqueue$ operations and
a single reader thread performing $\dequeue$ operations.
A suitable rely condition for $\enqueue$ is that 
elements are only ever removed from the front of the queue,
i.e.~the queue after any interference is a suffix of the queue before.
The rely condition for $\dequeue$ is that the interference 
only ever adds elements to the end of the queue
and hence the queue before the interference is a prefix of the queue after.
Note that the precondition of $\dequeue$ is stable under its rely condition
\cite{Wickerson10-TR,DBLP:conf/esop/WickersonDP10},
i.e.~$qu \neq [~] \land qu \prefix qu' \entails qu' \neq [~]$.
The guarantees of each operation match the rely of the other operation.
This version requires that there is just one reader thread and one writer thread
because the rely condition of $\dequeue$ assumes the queue can only be extended 
and hence it is not concurrently being dequeued by another thread,
and the rely condition of $\enqueue$ assumes the queue can only become a suffix of its previous state
and hence it is not concurrently being enqueued by another thread.

For sequential programs and in the original rely/guarantee approach
the postconditions of operations are considered end-to-end,
i.e.~they must hold between the states at the start and end of an operation invocation.
Such a postcondition is problematic in the context of a parallel thread modifying the queue,
for example, after an $\enqueue$ operation is initiated but before it can complete 
(or lock the data structure),
the reader thread may $\dequeue$ a value. 
If the writer then completes the $\enqueue$ without further interference, 
the end-to-end effect is that of both the $\dequeue$ and the $\enqueue$, not just the $\enqueue$.
An end-to-end postcondition is not suitable in this case.
The alternative is a specification whereby the postcondition holds for some ``atomic''
step between the start and end of the operation and the operation makes no changes
to the queue before or after that step, although other threads may.
Atomicity can be handled by using resources (see Sect.~\ref{S-resources}).
The environment may also interfere with operation preconditions.
For example, if multiple readers were allowed on a queue,
a precondition that the queue is non-empty may be invalidated
by a parallel $\dequeue$ operation that reads the last value from the queue thus making it empty.

\subsection{Resources}\label{S-resources}

\begin{figure}\normalsize
\figurerule
\begin{displaymath}
 \begin{array}{l}
  \Resource qu : \seq Val~~\Initially qu = [~] \\[1ex]
  \enqueue(v : Val) \sdefs \\~~~
  \begin{array}{l}
    (\rely{qu' \suffix qu}) \together ~~~\Comment{implies single writer} \\
    (\guar{qu \prefix qu'}) \together \Comment{implies the rely of \dequeue}  \\
    \with{qu}{\Spec{qu}{}{qu' = qu \cat [v]}}
  \end{array} \\[4ex]
  \dequeue() res : Val \sdefs \\~~~
  \begin{array}{l}
    (\rely{qu \prefix qu'}) \together ~~~\Comment{implies single reader} \\
    (\guar{qu' \suffix qu}) \together  ~\Comment{implies the rely of \enqueue} \\
    \Pre{qu \neq [~]} \Seq \Comment{stable under the rely condition} \\
    \with{qu}{\Spec{qu,res}{}{qu = [res'] \cat qu'}}
  \end{array}
 \end{array}
\end{displaymath}
\figurerule
\caption{Message queue with \dequeue\ and \enqueue\ operations}\label{F-queue}
\end{figure}

Early work of Hoare \cite{Hoare71g} introduced the idea of a resource and 
a $\With$ statement that provides access to the resource.
A \emph{resource} represents a shared data structure that is only accessible to a thread
within a command of the form,
\begin{eqnarray}
  \with{d}{c}
\end{eqnarray}
that ensures the resource $d$ is not modified by the environment while the thread is executing $c$.
A data structure $d$ is declared as a resource by a declaration of the form $\Resource d$
and, within the scope of the resource declaration, all uses of $d$ must be within a $\with{d}{...}$ statement.%
\footnote{A more general resource construct would allow a resource to encompass a set of variables
but for the examples here a resource will correspond to a single variable, 
so we identify the resource with the variable to simplify the presentation.}
It is required that the data structure of the resource is only accessed within $\With$ statements;
this may be checked syntactically.
The implementation is responsible for ensuring the data structure is not modified by the environment 
while executing within the $\With$ statement.
The $\With$ statement allows any number of stuttering steps before the body and 
finite stuttering after the body of the $\With$ is executed.
As entry to a $\With$ statement by one thread may block other threads wishing to gain access to the same resource, 
it is prudent to require that the bodies of $\With$ statements terminate;
that is required within this paper.
The data structure of the resource also often has a data-type invariant associated with it
that is established by its initialisation and maintained by each operation 
(see the example in Section~\ref{S-blocking}).

An operation whose body consists of a single $\With$ statement executes the $\With$ statement
atomically at some point of time between its invocation and completion.
Hence multiple operations that overlap in time will have their bodies executed in some sequence
(i.e.~their execution is linearized \cite{HW90})
and hence they will appear to behave equivalently to some sequential execution of the same operations,
i.e.~their execution is sequentially consistent.

\subsection{Rely/guarantee laws for resource access}

The concept of a resource may be combined with the rely/guarantee approach.
When a thread enters the body of a $\with{d}{c}$ statement,
the rely can be strengthened for the duration of $c$ with $d\,' = d$
and the guarantee weakened so that $d$ only need satisfy the guarantee over the complete operation,
not over every step.
The initial step of the refinement of operations specified via a $\With$ statement 
needs to ``move'' the rely and guarantee conditions into the body of the $\With$
but in the process the rely and guarantee conditions are transformed.
\begin{law}[rely-with]\label{law-rely-with}
\(
  ~~~~(\rely{r}) \together \with{d}{c} ~\refsto~ \with{d}{(\rely{r \land d\,'=d}) \together c}
\)
\end{law}
For the $\enqueue$ operation above (ignoring the guarantee for the moment) this law can be applied as follows.
\begin{displaymath}
 \begin{array}{l}
  (\rely{qu' \suffix qu}) \together \with{qu}{\Spec{qu}{}{qu' = qu \cat [v]}}
  \Refsto*[by \Law{rely-with}]
  \with{qu}{(\rely{qu' \suffix qu \land qu' = qu}) \together \Spec{qu}{}{qu' = qu \cat [v]}}
  \Equals*[as $qu' = qu ~\entails~ qu' \suffix qu$]
  \with{qu}{(\rely{qu' = qu}) \together \Spec{qu}{}{qu' = qu \cat [v]}}
 \end{array}
\end{displaymath}

A guarantee condition surrounding a $\With$ may be weakened so that it only has
to apply for the resource data structure over the body of the $\With$ command.
It is assumed that the guarantee is of the form $g_d \land g_x$,
where $d$ is the only shared variable $g_d$ refers to, and $g_x$ does not refer to $d$.
The weakened guarantee $g_x$ is retained to handle references to
variables other than $d$ within the guarantee.
The specification $\Spec{}{}{g_d}$ requires $g_d$ to hold end-to-end over the body of the $\With$.
The lack of a frame allows any variables to be modified but when it is combined with $c$,
any frame of $c$ will apply to their weak conjunction.
\begin{law}[guar-with]\label{law-guar-with}
\(
  ~~~~(\guar{g_d \land g_x}) \together \with{d}{c} ~\refsto~ 
      \with{d}{(\guar{g_x}) \together \Spec{}{}{g_d} \together c}
\)
\end{law}
For the $\enqueue$ operation above (ignoring the rely for the moment) this law can be applied as follows.
\begin{displaymath}
 \begin{array}{l}
  (\guar{qu \prefix qu'}) \together \with{qu}{\Spec{qu}{}{qu' = qu \cat [v]}}
  \Refsto*[by \Law{guar-with} with $g_d \sdefs qu \prefix qu'$ and $g_x \sdefs \True$]
  \with{qu}{(\guar{\True}) \together \Spec{}{}{qu \prefix qu'} \together \Spec{qu}{}{qu' = qu \cat [v]}}
  \Equals*[as $(\guar{\True})$ requires no guarantee and $\Spec{}{}{q_1} \together \Spec{x}{}{q_2} = \Spec{x}{}{q_1 \land q_2}$]
  \with{qu}{\Spec{qu}{}{qu \prefix qu' \land qu' = qu \cat [v]}}
  \Equals*[as {$qu' = qu \cat [v] ~\entails~ qu \prefix qu'$}]
  \with{qu}{\Spec{qu}{}{qu' = qu \cat [v]}}
 \end{array}
\end{displaymath}
Combining the above applications of \Law{rely-with} and \Law{guar-with},
the following refinement of the $\enqueue$ operation holds.
\begin{displaymath}
 \begin{array}{l}
    (\rely{qu' \suffix qu}) \together (\guar{qu \prefix qu'}) \together \with{qu}{\Spec{qu}{}{qu' = qu \cat [v]}}
   \Refsto*[using \Law{guar-with} and \Law{rely-with}]
    \with{qu}{(\rely{qu' = qu}) \together \Spec{qu}{}{qu' = qu \cat [v]}}
 \end{array}
\end{displaymath}
Note that in the body of the $\With$ there is no explicit guarantee and 
the rely condition assumes that $qu$ is not modified by the environment
and hence the refinement of the body of the $\With$ is effectively a sequential refinement
(as one would expect).

\pagebreak[3]

\section{Specifying operations that may block}\label{S-blocking}

\begin{figure}\normalsize
\figurerule
\begin{displaymath}
 \begin{array}{l}
  \Resource qu : \seq Val~~\Invariant~\# qu \leq N~~\Initially qu = [~] \\[1ex]
  \enqueue(v : Val) \sdefs \\~~~
  \begin{array}{l}
    (\rely{qu' \suffix qu}) \together \Comment{implies single writer} \\
    (\guar{qu \prefix qu'}) \together \Comment{implies the rely of \dequeue} \\
    \withawait{qu}{\# qu < N}{\Comment{stable under the rely condition} \\~~~
      \Pre{\# qu < N} \Seq \Spec{qu}{}{qu' = qu \cat [v]}
    }
  \end{array} \\[5ex]
  \dequeue() res : Val \sdefs \\~~~
  \begin{array}{l}
    (\rely{qu \prefix qu'}) \together \Comment{implies single reader} \\
    (\guar{qu' \suffix qu}) \together \Comment{implies the rely of \enqueue} \\
    \withawait{qu}{qu \neq [~]}{\Comment{stable under the rely condition} \\~~~
      \Pre{qu \neq [~]} \Seq \Spec{qu,res}{}{qu = [res'] \cat qu'}
    }
  \end{array}
 \end{array}
\end{displaymath}
\figurerule
\caption{Message queue with blocking \dequeue\ and \enqueue\ operations}\label{F-blocking-queue}
\end{figure}

\subsection{Blocking using an explicit await condition}\label{S-blocking-await}

Consider the example in Fig.~\ref{F-blocking-queue} of a message queue with a bounded capacity of $N$ messages.
It has a $\enqueue$ operation that waits until the queue is not full and
appends a value to the tail of the queue, and
a $\dequeue$ operation that waits until there is a message in the queue 
and returns the head of the queue, removing it in the process.
The queue has a data-type invariant that its size is bounded by $N$.
The initialisation establishes the invariant (assuming $N$ is a positive integer)
and each operation on the queue can assume the invariant when it starts
and must be re-establish the invariant when it terminates.

A common approach to specifying potentially blocking operations is to use an $\Await$ statement.
The form used here is combined with a $\With$ statement.
The statement $\withawait{d}{b}{c}$ waits until condition $b$ holds and then executes $c$.
The resource $d$ is attained each time $b$ is evaluated, 
and retained for the execution of $c$ if $b$ is true;
if $b$ is false the resource is released and the await retries. 
For both the $\enqueue$ and $\dequeue$ operations the await conditions are stable under the rely condition,
i.e.~if the await condition $b$ holds before a step that satisfies the rely condition $r$,
then $b$ holds after the step.
Hence once the operations are enabled, they remain enabled until they occur.

\subsection{Blocking using temporal logic termination conditions}\label{S-blocking-assume}

\begin{figure}\normalsize
\figurerule
\begin{displaymath}
 \begin{array}{l}
  \Resource qu : \seq Val~~\Invariant~\# qu \leq N~~\Initially qu = [~] \\[1ex]
  \enqueue(v : Val) \sdefs \\~~~
  \begin{array}{l}
    \Terminate \eventually(\# qu < N) \together {}\\~~~
     \begin{array}{l}
      (\rely{qu' \suffix qu}) \together \Comment{implies single writer} \\
      (\guar{qu \prefix qu'}) \together \Comment{implies the rely of \dequeue} \\
      \with{qu}{\Spec{qu}{}{qu' = qu \cat [v]}}
    \end{array}
  \end{array} \\[5ex]
  \dequeue() res : Val \sdefs \\~~~
  \begin{array}{l}
    \Terminate \eventually(qu \neq [~]) \together {}\\~~~
     \begin{array}{l}
      (\rely{qu \prefix qu'}) \together \Comment{implies single reader} \\
      (\guar{qu' \suffix qu}) \together \Comment{implies the rely of \enqueue} \\
      \with{qu}{\Spec{qu,res}{}{qu = [res'] \cat qu'}}
    \end{array}
  \end{array}
 \end{array}
\end{displaymath}
\figurerule
\caption{Message queue with conditions to ensure termination}\label{F-assume-queue}
\end{figure}

The specifications of $\dequeue$ and $\enqueue$ in Fig.~\ref{F-blocking-queue}, by including $\Await$ statements,
allow non-terminating behaviour in the cases where the await condition never becomes true,
e.g.\ a $\dequeue$ will wait forever if the queue remains empty because no writes are performed.
However, 
$\dequeue$ terminates if the queue is eventually non-empty
and
$\enqueue$ terminates if the queue is eventually non-full.
That leads to an alternative specification using temporal logic termination conditions in Fig.~\ref{F-assume-queue}.
To accommodate the termination conditions, a command of the form,
\begin{eqnarray}
  \Terminate t  & \sdefs & \Term \nondet (\Encode \lnot t)
\end{eqnarray}
is introduced, in which 
the command $\Term$ only allows any terminating behaviour 
and for a temporal logic formula $t$,
the command $\Encode t$ allows just those traces that satisfy $t$.%
\footnote{Such a command can be straightforwardly defined in the semantics for rely/guarantee
concurrency presented in \cite{DaSMfaWSLwC}.}
If the temporal logic formula $t$ holds, $\Terminate t$ must terminate,
but if $t$ does not hold termination is not required, but is allowed.
Neither of the specifications of $\enqueue$ and $\dequeue$ in Fig.~\ref{F-assume-queue},
contain the explicit $\Await$s used in Fig.~\ref{F-blocking-queue}.
The postcondition of the $\dequeue$ is unsatisfiable if the queue always remains empty because 
\begin{eqnarray*}
[~] = [res'] \cat qu' & \equiv & \False~.
\end{eqnarray*}
However, if the condition $\eventually(qu \neq [~])$ holds
(i.e.~the queue is eventually non-empty), 
the postcondition eventually becomes feasible from the state in which the queue is non-empty.
Note that $qu \neq [~]$ is stable under the rely condition and
so it will not be falsified by the environment
and hence the termination condition is equivalent to $\eventually \always(qu \neq [~])$ in this case.

For the $\dequeue$ operation, the negation of $\eventually(qu \neq [~])$ is
$\always(qu = [~])$, i.e.~the queue is always empty.
If $\always(qu = [~])$ the $\dequeue$ operation is not required to terminate.
In addition, if in every state $qu = [~]$,
the postcondition of $\dequeue$ is unsatisfiable because $[~] = [res'] \cat [~]$ is false,
and hence the terminating behaviour of $\dequeue$ is infeasible,
and therefore the only possible behaviour if $\dequeue$ is to not terminate.

More subtly, any finite prefix of a trace of a $\dequeue$ operation for which $qu = [~]$ in every state 
cannot have satisfied the postcondition of $\dequeue$
and hence cannot have terminated.
However, it is still possible that another thread may execute a $\enqueue$ at some later time
establishing $qu \neq [~]$ and allowing the $\dequeue$ to terminate.
For the finite prefix of the trace, the behaviour of the $\dequeue$ must correspond to 
the stuttering allowed by the $\With$ statement before it enters its body.
Hence for every finite prefix of a trace of $\dequeue$ in which $qu = [~]$ in every state,
the operation performs only finite stuttering steps
and, further, nontermination is allowed if $\always(qu = [~])$.
Hence the behaviours allowed by the specifications in Fig.~\ref{F-assume-queue} are 
actually equivalent to those of the specifications in Fig.~\ref{F-blocking-queue}.

For the queue example, if multiple readers and writers are allowed,
the specifications need to be adapted.
Firstly, the rely conditions of the $\enqueue$ and $\dequeue$ operations need to
be removed because the rely of $\enqueue$ is broken by a concurrent thread 
also performing $\enqueue$s, 
and similarly for the rely of $\dequeue$ if there are concurrent readers.
Secondly, the termination conditions need to be strengthened.
There are two strengthenings that correspond to weakly and strongly fair
interpretations of the await in specifications using $\Await$.

For weak fairness 
the termination condition of $\dequeue$ is strengthened to $\eventually \always(qu \neq [~])$,
that is, eventually the queue remains non-empty,
which implies that concurrent threads are not making the queue empty through $\dequeue$ operations.
For the equivalent specification using an $\Await$ rather than a $\Terminate$,
the $\Await$ can be given a weakly fair interpretation,
i.e.~if the guard of the await for $\dequeue$, $qu \neq [~]$, is eventually continually enabled,
the $\Await$ will succeed. 
The termination condition for $\enqueue$ is similar.

For strong fairness 
the termination condition of $\dequeue$ is the weaker $\always \eventually (qu \neq [~])$,
that is, it is always the case that eventually the queue is non-empty,
which implies that concurrent threads can repeatedly make the queue empty through $\dequeue$ operations
but if the queue becomes empty there must then be one of more $\enqueue$ operations 
that make it non-empty.
Because the termination condition is weaker, 
this version of $\dequeue$ must terminate for a wider range of behaviours of its environment
and hence requires a stronger implementation,
e.g.~one that strictly sequences access to the queue, perhaps using a ticket lock.

For the equivalent specification using an $\Await$ rather than a $\Terminate$,
the $\Await$ can be given a strongly fair interpretation,
i.e.~if the guard of the await for $\dequeue$, $qu \neq [~]$, is always eventually enabled
(but the queue may alternate between empty and non-empty),
the $\Await$ will succeed. 

An advantage of the specifications using $\Terminate$ clauses is 
that the condition under which an operation is guaranteed to terminate is made explicit,
unlike specifications using await statements 
where the semantics of the await needs to be changed.
This makes it clearer what a programmer using a component can rely on.
It also avoids the complication of having different interpretations of await guards
as used by Liang and Feng \cite{LiangFengPOPL18}.

\section{Transactional operations}\label{S-transactional}

Some implementations of operations are optimistic in that 
they complete most of the operation locally within a thread and 
then have a final commit phase that may fail if another operation has committed.
Such operations consist of a repeated failure behaviour 
(that does not change the shared data structure)
in the presence of interference
followed by a successful commit phase.
Of course, in the presence of repeated interference the successful commit may never occur.
There are many examples of lock-free algorithms that exhibit the above behaviour \cite{Scott13}
but to make the discussion concrete, 
we make use of Treiber's well-known concurrent stack data structure \cite{Treiber86}.

\subsection{Specification using explicit failure}\label{S-transactional-explicit}

\begin{figure}\normalsize
\figurerule
\begin{displaymath}
 \begin{array}{l}
  \Resource s : \seq Val~~\Initially s = [~] \\[2ex]
  push(v : Val) \sdefs \Pre{v \neq \Const{null}} \Seq \Om{push\_fail} \Seq push\_success(v) \\[0ex]
  \Where 
   \begin{array}[t]{l}
    push\_fail \sdefs \EnvAtomic{s' \neq s} \\
    push\_success(v : Val) \sdefs \with{s}{\Spec{s}{}{s' = [v] \cat s}} \together \Term
   \end{array}\\[4ex]
  pop() res : Val \sdefs \Om{pop\_fail} \Seq res:= pop\_success() \\[0ex]
  \Where 
   \begin{array}[t]{l}
    pop\_fail \sdefs \EnvAtomic{s' \neq s} \\
    pop\_success() res : Val \sdefs 
      \with{s}{\Spec{s,res}{}{(s \neq [~] \implies s = [res'] \cat s') \land \\(s = [~] \implies res' = \Const{null}}} \together \Term
   \end{array}
 \end{array}
\end{displaymath}
\figurerule
\caption{Stack with possibly failing push and pop operations}\label{F-stack}
\end{figure}

Treiber \cite{Treiber86} provided a non-blocking lock-free implementation of a stack
in which the push and pop operations may be concurrently executed by many threads.
However, an attempt to push or pop a value may fail due to interference from parallel
stack operations, and hence it may need to be retried until it succeeds.
In the worse case, if every time an operation is tried it fails due to interference,
that invocation of  operation may never terminate.

Fig.~\ref{F-stack} gives a specification of a stack with $push$ and $pop$ operations that may fail and need to retry,
possibly indefinitely.
If the stack is empty, $pop$ returns the special value $\Const{null}$,
which may not be pushed onto the stack.%
\footnote{In the context of concurrent stack operations, 
the $pop$ operation must well defined when the stack is empty;
note that even if an $is\_empty$ operation returned false,
concurrent interference may empty the stack before a subsequent $pop$ operation.}
The $push\_fail$ operation may be executed any number of times
but each time it is executed the environment makes a step that changes the stack $s$.
If from some point of time the environment never changes $s$, 
then the $push\_fail$ becomes infeasible and the operation must perform $push\_success$
which pushes the value on the stack and terminates;
the command $\Term$ only allows any terminating computation and hence
when conjoined with a specification restricts it to just its terminating behaviours. 
The command $\EnvAtomic{s' \neq s}$ corresponds to the environment performing a step that modifies $s$;
it may also perform a finite number of stuttering program steps 
(i.e.\ steps that do not change observable variables).
If the environment performs a step modifying $s$, $\EnvAtomic{s' \neq s}$ terminates
but if not, it becomes infeasible
and termination of the iteration is forced, so that the $push\_success$ alternative is taken.
The definition of $pop$ is similar.

\subsection{Specification using temporal logic termination conditions}\label{S-transactional-assume}

Note that if the stack $s$ is never changed by the environment,
the behaviour of the $push$ (or $pop$) operation reduces to just its successful behaviour.
More subtly, if the environment eventually stops changing $s$,
then there can only be a finite number of failure iterations before the operation succeeds.
This latter condition can be converted into a temporal logic termination condition 
$\eventually \always_{\epsilon}(s' = s)$,
i.e.~eventually all environment steps do not change the value of the stack,
which leads to the specification given in Fig.~\ref{F-stack-assume}.
An extended form of temporal logic is used here that distinguishes program and environment steps
and allows one to specify a constraint on a step in the form of a relation,
in this case $s'=s$.
If parallel activity on the stack eventually quiesces, 
Treiber's push and pop operations are guaranteed to terminate,
and hence the specifications with the quiescence termination conditions do not need to include the failure possibilities.

\begin{figure}\normalsize
\figurerule
\begin{displaymath}
 \begin{array}{l}
  \Resource s : \seq Val~~\Initially s = [~] \\[2ex]
  push(v : Val) \sdefs \Pre{v \neq \Const{null}} \Seq 
      \Terminate \eventually \always_\epsilon(s' = s) \together \with{s}{\Spec{s}{}{s' = [v] \cat s}} \\[2ex]
  pop() res : Val \sdefs \Terminate \eventually \always_\epsilon(s' = s) \together \with{s}{\Spec{s,res}{}{(s \neq [~] \implies s = [res'] \cat s') \land \\(s = [~] \implies res' = \Const{null}}}
 \end{array}
\end{displaymath}
\figurerule
\caption{Stack with conditions to ensure termination}\label{F-stack-assume}
\end{figure}

The negation of the termination condition is $\always \eventually_\epsilon(s' \neq s)$,
i.e.~from every state there is eventually an environment step that modifies $s$.
However --unlike for the blocking queue-- that does not make the postcondition of either operation unsatisfiable
and hence if the negation of the termination condition holds, 
each operation may either terminate satisfying its postcondition or never terminate.
In a similar manner to the blocking queue, 
for a finite trace for which an operation has not yet satisfied its postcondition,
it is still possible to extend the trace so that the postcondition is satisfied,
and hence the only allowable behaviour of the operation for a finite trace 
that has not yet satisfied its postcondition is finite stuttering.
Hence if the termination condition is not satisfied an operation may either 
terminate successfully satisfying its postcondition
or
fail to terminate but only ever perform stuttering steps,
i.e.~it never modifies $s$.
Hence the specifications in Fig.~\ref{F-stack-assume} are equivalent to those in Fig.~\ref{F-stack}.

If the termination conditions on the stack operations are replaced by $true$,
the operations must always terminate, 
even under interference from other threads performing $push$ and $pop$ operations.
That gives strictly stronger specifications because their termination conditions are weaker.
An implementation might then be required to make use of a lock that sequentialises access to the stack
in the order in which the lock is requested (such as a ticket lock) in order to ensure termination.

\section{Conclusions}\label{S-conclusions}

To specify concurrent program components one needs to be able to address issues such as
operation atomicity, 
operations blocking on conditions or locks, and 
transactional operations that may fail and need to be retried.
Hoare's resource concept provides a notion of atomicity with respect to a resource.
A contribution of this paper is to examine its interaction with rely and guarantee conditions
in order to enable the initial refinement step of Hoare's $\With$ statements to code.
Brookes \cite{BrookesSepLogic07} also makes use of the concept of a resource in concurrent separation logic.
He generalises Hoare's concept to handle the heap as well as variables.

The specifications of operations using $\With$ statements do not dictate whether they
are refined to implementations using locks or to non-blocking implementations
or even a programming language that supports $\With$ statements.
One issue not addressed here is that operations requiring multiple resources,
e.g.~an operation that needs to perform operations on two separate resources and 
needs to be considered atomic as a whole.
In this case the $\With$ statement needs to allow for multiple resources,
and if locking is used in the implementation of the operations,
to avoid deadlock,
the locking has to ensure that resources are locked in the same order.
Note that $\With$ statements may be nested but they do not allow the complete flexibility of locks.
For example, algorithms that lock one node in a list and then lock the next node
before releasing the first node do not match a nesting structure.

Operations that block waiting for some condition have the potential for non-terminating behaviour.
That has been addressed via two ways of specifying such operations:
a form that makes use of an $\Await$ construct that blocks until its guard holds,
and
an implicit form that includes a condition under which termination is required.

Non-blocking algorithms can provide more efficient solutions for managing shared data structures
than using locks, but some algorithms have the issue that, under interference,
they may fail and need to be retried.
In the worst case an operation may be continually thwarted and never get a chance to complete
and hence its specification needs to either allow for that possible behaviour
or provide conditions under which it will terminate successfully,
e.g.~that the interference quiesces so that it can complete.

The approach taken in this paper is to indicate some directions for devising specifications
for concurrent program components.
In doing so we have shown that specifications with explicit await clauses can be
expressed in a more abstract form with a temporal logic formula giving the condition
under which termination is guaranteed.
Such specifications have greater expressive power than those using explicit await constructs
because temporal logic formulae allow termination conditions 
that cannot be expressed as await conditions,
for example, if the blocking queue allows multiple readers 
and hence its await condition was no longer stable,
an alternative termination condition of $\always \eventually(qu \neq [~])$
would require an implementation to guarantee the termination of each \dequeue\ operation
under interference from other \dequeue s,
provided a writer was also actively appending values to the queue.
An implementation of $\dequeue$ might, for example, use a lock that sequentialises access
in the order in which the lock is requested (such as a ticket lock)
in order to ensure termination.

Liang and Feng \cite{LiangFengPOPL18} have addressed handling progress conditions
for blocking operations (which they refer to as partial methods).
Their approach makes use of await statements 
but they give four different semantic interpretations to await statements 
depending on whether one requires the operations to be starvation free or deadlock free,
and depending on whether the enabling conditions are treated as weakly or strongly fair.
The weakly and strongly fair interpretations correspond to different termination conditions for operations.
The explicit use of a temporal logic termination condition differentiates these two cases
while avoiding the issues involved with giving different semantic interpretations to the await construct.

At this stage our treatment has not been fully formalised and
further work is required to support refinement of such specifications to code.

\paragraph{Acknowledgements.}

Thanks are due to
Robert Colvin,
Cliff Jones,
Larissa Meinicke,
Patrick Meiring,
and
Kirsten Winter,
and the anonymous reviewers
for feedback on the ideas presented here.
This research was supported
Australian Research Council Discovery Grant DP130102901.

\bibliographystyle{eptcs}
\bibliography{ms}

\begin{thebibliography}{10}
\providecommand{\bibitemdeclare}[2]{}
\providecommand{\surnamestart}{}
\providecommand{\surnameend}{}
\providecommand{\urlprefix}{Available at }
\providecommand{\url}[1]{\texttt{#1}}
\providecommand{\href}[2]{\texttt{#2}}
\providecommand{\urlalt}[2]{\href{#1}{#2}}
\providecommand{\doi}[1]{doi:\urlalt{http://dx.doi.org/#1}{#1}}
\providecommand{\bibinfo}[2]{#2}

\bibitemdeclare{article}{BrookesSepLogic07}
\bibitem{BrookesSepLogic07}
\bibinfo{author}{S.~\surnamestart Brookes\surnameend} (\bibinfo{year}{2007}):
  \emph{\bibinfo{title}{A semantics for concurrent separation logic}}.
\newblock {\sl \bibinfo{journal}{Theoretical Computer Science}}
  \bibinfo{volume}{375}(\bibinfo{number}{1--3}), pp. \bibinfo{pages}{227--270},
  \doi{10.1016/j.tcs.2006.12.034}.

\bibitemdeclare{article}{DaSMfaWSLwC}
\bibitem{DaSMfaWSLwC}
\bibinfo{author}{R.~J. \surnamestart Colvin\surnameend}, \bibinfo{author}{I.~J.
  \surnamestart Hayes\surnameend} \& \bibinfo{author}{L.~A. \surnamestart
  Meinicke\surnameend} (\bibinfo{year}{2016}): \emph{\bibinfo{title}{Designing
  a semantic model for a wide-spectrum language with concurrency}}.
\newblock {\sl \bibinfo{journal}{Formal Aspects of Computing}}
  \bibinfo{volume}{29}, pp. \bibinfo{pages}{853--875},
  \doi{10.1007/s00165-017-0416-4}.

\bibitemdeclare{inproceedings}{Floyd67}
\bibitem{Floyd67}
\bibinfo{author}{R.~W. \surnamestart Floyd\surnameend} (\bibinfo{year}{1967}):
  \emph{\bibinfo{title}{Assigning meanings to programs}}.
\newblock In: {\sl \bibinfo{booktitle}{Proceedings of Symposia in Applied
  Mathematics: Math. Aspects of Comput. Sci.}}, \bibinfo{volume}{19}, pp.
  \bibinfo{pages}{19--32}, \doi{10.1090/psapm/019/0235771}.

\bibitemdeclare{article}{AFfGRGRACP}
\bibitem{AFfGRGRACP}
\bibinfo{author}{I.~J. \surnamestart Hayes\surnameend} (\bibinfo{year}{2016}):
  \emph{\bibinfo{title}{Generalised rely-guarantee concurrency: An algebraic
  foundation}}.
\newblock {\sl \bibinfo{journal}{Formal Aspects of Computing}}
  \bibinfo{volume}{28}(\bibinfo{number}{6}), pp. \bibinfo{pages}{1057--1078},
  \doi{10.1007/s00165-016-0384-0}.

\bibitemdeclare{inproceedings}{FM2016atomicSteps}
\bibitem{FM2016atomicSteps}
\bibinfo{author}{I.~J. \surnamestart Hayes\surnameend}, \bibinfo{author}{R.~J.
  \surnamestart Colvin\surnameend}, \bibinfo{author}{L.~A. \surnamestart
  Meinicke\surnameend}, \bibinfo{author}{K.~\surnamestart Winter\surnameend} \&
  \bibinfo{author}{A.~\surnamestart Velykis\surnameend} (\bibinfo{year}{2016}):
  \emph{\bibinfo{title}{An algebra of synchronous atomic steps}}.
\newblock In \bibinfo{editor}{J.~\surnamestart Fitzgerald\surnameend},
  \bibinfo{editor}{C.~\surnamestart Heitmeyer\surnameend},
  \bibinfo{editor}{S.~\surnamestart Gnesi\surnameend} \&
  \bibinfo{editor}{A.~\surnamestart Philippou\surnameend}, editors: {\sl
  \bibinfo{booktitle}{FM 2016: Formal Methods: 21st International Symposium,
  Proceedings}}, {\sl \bibinfo{series}{LNCS}} \bibinfo{volume}{9995},
  \bibinfo{publisher}{Springer International Publishing},
  \bibinfo{address}{Cham}, pp. \bibinfo{pages}{352--369},
  \doi{10.1007/978-3-319-48989-6_22}.

\bibitemdeclare{techreport}{HayesJonesColvin14TR}
\bibitem{HayesJonesColvin14TR}
\bibinfo{author}{I.~J. \surnamestart Hayes\surnameend}, \bibinfo{author}{C.~B.
  \surnamestart Jones\surnameend} \& \bibinfo{author}{R.~J. \surnamestart
  Colvin\surnameend} (\bibinfo{year}{2014}): \emph{\bibinfo{title}{Laws and
  semantics for rely-guarantee refinement}}.
\newblock \bibinfo{type}{Technical Report} \bibinfo{number}{CS-TR-1425},
  \bibinfo{institution}{Newcastle University}.

\bibitemdeclare{article}{FMJournalAtomicSteps}
\bibitem{FMJournalAtomicSteps}
\bibinfo{author}{Ian~J. \surnamestart Hayes\surnameend},
  \bibinfo{author}{Larissa~A. \surnamestart Meinicke\surnameend},
  \bibinfo{author}{Kirsten \surnamestart Winter\surnameend} \&
  \bibinfo{author}{Robert~J. \surnamestart Colvin\surnameend}
  (\bibinfo{year}{2018}): \emph{\bibinfo{title}{A synchronous program algebra:
  a basis for reasoning about shared-memory and event-based concurrency}}.
\newblock {\sl \bibinfo{journal}{Formal Aspects of Computing}},
  \doi{10.1007/s00165-018-0464-4}.
\newblock \bibinfo{note}{Online 6 August 2018}.

\bibitemdeclare{article}{HW90}
\bibitem{HW90}
\bibinfo{author}{Maurice \surnamestart Herlihy\surnameend} \&
  \bibinfo{author}{Jeannette~M. \surnamestart Wing\surnameend}
  (\bibinfo{year}{1990}): \emph{\bibinfo{title}{Linearizability: A Correctness
  Condition for Concurrent Objects.}}
\newblock {\sl \bibinfo{journal}{ACM Trans. Program. Lang. Syst.}}
  \bibinfo{volume}{12}(\bibinfo{number}{3}), pp. \bibinfo{pages}{463--492},
  \doi{10.1145/78969.78972}.

\bibitemdeclare{article}{Hoare69a}
\bibitem{Hoare69a}
\bibinfo{author}{C.~A.~R. \surnamestart Hoare\surnameend}
  (\bibinfo{year}{1969}): \emph{\bibinfo{title}{An Axiomatic Basis for Computer
  Programming}}.
\newblock {\sl \bibinfo{journal}{Communications of the ACM}}
  \bibinfo{volume}{12}(\bibinfo{number}{10}), pp. \bibinfo{pages}{576--580,
  583}, \doi{10.1145/363235.363259}.

\bibitemdeclare{inproceedings}{Hoare71g}
\bibitem{Hoare71g}
\bibinfo{author}{C.~A.~R. \surnamestart Hoare\surnameend}
  (\bibinfo{year}{1972}): \emph{\bibinfo{title}{Towards a Theory of Parallel
  Programming}}.
\newblock In: {\sl \bibinfo{booktitle}{Operating System Techniques}},
  \bibinfo{publisher}{Academic Press}, pp. \bibinfo{pages}{61--71}.

\bibitemdeclare{article}{Hoare75}
\bibitem{Hoare75}
\bibinfo{author}{C.~A.~R. \surnamestart Hoare\surnameend}
  (\bibinfo{year}{1975}): \emph{\bibinfo{title}{Parallel programming: an
  axiomatic approach}}.
\newblock {\sl \bibinfo{journal}{Computer Languages}}
  \bibinfo{volume}{1}(\bibinfo{number}{2}), pp. \bibinfo{pages}{151--160},
  \doi{10.1016/0096-0551(75)90014-4}.

\bibitemdeclare{phdthesis}{Jones81d}
\bibitem{Jones81d}
\bibinfo{author}{C.~B. \surnamestart Jones\surnameend} (\bibinfo{year}{1981}):
  \emph{\bibinfo{title}{Development Methods for Computer Programs including a
  Notion of Interference}}.
\newblock Ph.D. thesis, \bibinfo{school}{Oxford University}.
\newblock \bibinfo{note}{Available as: Oxford University Computing Laboratory
  (now Computer Science) Technical Monograph PRG-25}.

\bibitemdeclare{inproceedings}{Jones83a}
\bibitem{Jones83a}
\bibinfo{author}{C.~B. \surnamestart Jones\surnameend} (\bibinfo{year}{1983}):
  \emph{\bibinfo{title}{Specification and Design of (Parallel) Programs}}.
\newblock In: {\sl \bibinfo{booktitle}{Proceedings of IFIP'83}},
  \bibinfo{publisher}{North-Holland}, pp. \bibinfo{pages}{321--332}.

\bibitemdeclare{article}{Jones83b}
\bibitem{Jones83b}
\bibinfo{author}{C.~B. \surnamestart Jones\surnameend} (\bibinfo{year}{1983}):
  \emph{\bibinfo{title}{Tentative Steps Toward a Development Method for
  Interfering Programs}}.
\newblock {\sl \bibinfo{journal}{ACM ToPLaS}}
  \bibinfo{volume}{5}(\bibinfo{number}{4}), pp. \bibinfo{pages}{596--619},
  \doi{10.1145/69575.69577}.

\bibitemdeclare{article}{LiangFengPOPL18}
\bibitem{LiangFengPOPL18}
\bibinfo{author}{Hongjin \surnamestart Liang\surnameend} \&
  \bibinfo{author}{Xinyu \surnamestart Feng\surnameend} (\bibinfo{year}{2018}):
  \emph{\bibinfo{title}{Progress of Concurrent Objects with Partial Methods}}.
\newblock {\sl \bibinfo{journal}{Proc. ACM Program. Lang.}}
  \bibinfo{volume}{2}(\bibinfo{number}{POPL}), pp.
  \bibinfo{pages}{20:1--20:31}, \doi{10.1145/3158108}.

\bibitemdeclare{article}{TSS}
\bibitem{TSS}
\bibinfo{author}{C.~C. \surnamestart Morgan\surnameend} (\bibinfo{year}{1988}):
  \emph{\bibinfo{title}{The Specification Statement}}.
\newblock {\sl \bibinfo{journal}{ACM Trans.\ Prog.\ Lang.\ and Sys.}}
  \bibinfo{volume}{10}(\bibinfo{number}{3}), pp. \bibinfo{pages}{403--419},
  \doi{10.1145/44501.44503}.

\bibitemdeclare{book}{Morgan94}
\bibitem{Morgan94}
\bibinfo{author}{C.~C. \surnamestart Morgan\surnameend} (\bibinfo{year}{1994}):
  \emph{\bibinfo{title}{Programming from Specifications}},
  \bibinfo{edition}{second} edition.
\newblock \bibinfo{publisher}{Prentice Hall}.

\bibitemdeclare{article}{OHearn07}
\bibitem{OHearn07}
\bibinfo{author}{P.~W. \surnamestart O'Hearn\surnameend}
  (\bibinfo{year}{2007}): \emph{\bibinfo{title}{Resources, Concurrency and
  Local Reasoning}}.
\newblock {\sl \bibinfo{journal}{Theoretical Computer Science}}
  \bibinfo{volume}{375}(\bibinfo{number}{1-3}), pp. \bibinfo{pages}{271--307},
  \doi{10.1016/j.tcs.2006.12.035}.

\bibitemdeclare{phdthesis}{Owicki75}
\bibitem{Owicki75}
\bibinfo{author}{S.~\surnamestart Owicki\surnameend} (\bibinfo{year}{1975}):
  \emph{\bibinfo{title}{Axiomatic Proof Techniques for Parallel Programs}}.
\newblock Ph.D. thesis, \bibinfo{school}{Department of Computer Science,
  Cornell University}.

\bibitemdeclare{article}{OwickiGries76}
\bibitem{OwickiGries76}
\bibinfo{author}{S.~S. \surnamestart Owicki\surnameend} \&
  \bibinfo{author}{D.~\surnamestart Gries\surnameend} (\bibinfo{year}{1976}):
  \emph{\bibinfo{title}{An axiomatic proof technique for parallel programs
  {I}}}.
\newblock {\sl \bibinfo{journal}{Acta Informatica}}
  \bibinfo{volume}{6}(\bibinfo{number}{4}), pp. \bibinfo{pages}{319--340},
  \doi{10.1007/BF00268134}.

\bibitemdeclare{article}{OwickiGries-CACM-76}
\bibitem{OwickiGries-CACM-76}
\bibinfo{author}{Susan \surnamestart Owicki\surnameend} \&
  \bibinfo{author}{David \surnamestart Gries\surnameend}
  (\bibinfo{year}{1976}): \emph{\bibinfo{title}{Verifying Properties of
  Parallel Programs: An Axiomatic Approach}}.
\newblock {\sl \bibinfo{journal}{Commun. ACM}}
  \bibinfo{volume}{19}(\bibinfo{number}{5}), pp. \bibinfo{pages}{279--285},
  \doi{10.1145/360051.360224}.

\bibitemdeclare{book}{Scott13}
\bibitem{Scott13}
\bibinfo{author}{Michael~L. \surnamestart Scott\surnameend}
  (\bibinfo{year}{2013}): \emph{\bibinfo{title}{Shared-Memory
  Synchronization}}.
\newblock \bibinfo{publisher}{Morgan \& Claypool Publishers}.

\bibitemdeclare{techreport}{Treiber86}
\bibitem{Treiber86}
\bibinfo{author}{R.~K. \surnamestart Treiber\surnameend}
  (\bibinfo{year}{1986}): \emph{\bibinfo{title}{Systems Programming: Coping
  with Parallelism.}}
\newblock \bibinfo{type}{Technical Report} \bibinfo{number}{RJ 5118},
  \bibinfo{institution}{IBM Almaden Research Center}.

\bibitemdeclare{techreport}{Wickerson10-TR}
\bibitem{Wickerson10-TR}
\bibinfo{author}{J.~\surnamestart Wickerson\surnameend},
  \bibinfo{author}{M.~\surnamestart Dodds\surnameend} \& \bibinfo{author}{M.~J.
  \surnamestart Parkinson\surnameend} (\bibinfo{year}{2010}):
  \emph{\bibinfo{title}{Explicit stabilisation for modular rely-guarantee
  reasoning}}.
\newblock \bibinfo{type}{Technical Report} \bibinfo{number}{774},
  \bibinfo{institution}{Computer Laboratory, University of Cambridge}.

\bibitemdeclare{inproceedings}{DBLP:conf/esop/WickersonDP10}
\bibitem{DBLP:conf/esop/WickersonDP10}
\bibinfo{author}{J.~\surnamestart Wickerson\surnameend},
  \bibinfo{author}{M.~\surnamestart Dodds\surnameend} \& \bibinfo{author}{M.~J.
  \surnamestart Parkinson\surnameend} (\bibinfo{year}{2010}):
  \emph{\bibinfo{title}{Explicit Stabilisation for Modular Rely-Guarantee
  Reasoning}}.
\newblock In \bibinfo{editor}{A.~D. \surnamestart Gordon\surnameend}, editor:
  {\sl \bibinfo{booktitle}{ESOP}}, {\sl \bibinfo{series}{LNCS}}
  \bibinfo{volume}{6012}, \bibinfo{publisher}{Springer}, pp.
  \bibinfo{pages}{610--629}.
\newblock \urlprefix\url{http://dx.doi.org/10.1007/978-3-642-11957-6_32}.

\end{thebibliography}

\end{document}